\documentstyle[11pt,newpasp,twoside,epsf]{article}
\begin{document}
\title{Variable Star Monitoring in Local Group Dwarf Irregular Galaxies}
 \author{Jan Snigula, Claus G\"ossl, Ulrich Hopp, Heinz Barwig}
\affil{Universit\"ats-Sternwarte M\"unchen, Scheiner Stra{\ss}e 1,
      D 81679 M\"unchen, Germany}

\begin{abstract}
  Dwarf galaxies in the local group provide a unique astrophysical
  laboratory. Despite their proximity some of these systems still lack
  a reliable distance determination as well as studies of their
  stellar content and star formation history. We present first results
  of our survey of variable stars in a sample of six local group dwarf
  irregular galaxies. Taking the Leo A dwarf galaxy as an example we
  describe observational strategies and data reduction. We discuss the
  lightcurves of two newly found $\delta$ Cephei stars and place them
  into the context of a previously derived P-L relation. Finally we
  discuss the LPV content of Leo~A.
\end{abstract}

\section{Introduction}

A magnitude limited complete census of variable stars in nearby dwarf
galaxies allows important contributions to the star formation history
of these systems. Measurements of some variable stars can supply
improved distance determinations for the host galaxies, others will
provide important constraints for the population analysis. Different
classes of variables can further improve the understanding of the star
formation history of these system, functioning as tracers of star
formation during different epochs. We expect the data set of our long
term monitoring program to be especially well suited to study the
contents of red long-period variables and to re-investigate the
paucity of Cepheids with $P>10$ days as reported by Sandage \& Carlson
(1985).

\section{Observations and data reduction}

We selected a sample of six local group dwarf irregular galaxies which
are visible with the 0.8~m telescope of our institute at Mt.
Wendelstein. The names and additional data from the literature
compilation by Mateo (1998) are shown in Table 1.

\begin{table}[t]
  \caption{Names, variable star counts, absolute $B$-Band brightness in mag,
    and current distance estimation in kpc for the dwarf galaxies observed in
    our project. The data are taken from the literature compilation by
    Mateo (1995). For Leo A the data are from the work of Dolphin
    et. al (2002) and from this work.}

  \begin{tabular}{ll|ccccc}
    \tableline
    \tableline
    & & RR Lyr & $\delta$ Cep & LPV & $M_B$ & distance\\\tableline
    LGS 3 & (Pisces)       & -- & -- & -- & -9.9 & $810 \pm 60$\\
    UGCA 92 & (EGB 0427+63)  & -- & -- & -- & -11.6 & $1300 \pm 700$\\
    DDO 69 & (Leo A) &        8 & 66 \it{+1}$^1$ &
    \it{16}$^1$ & -11.3 & $690 \pm 100$\\
    DDO 155 & (GR 8)        &  -- & 1? & 5? & -11.2 & $1590 \pm 600$\\    
    DDO 210 & (Aquarius)    &  -- & 0 & -- & -9.9 & $800 \pm 250$\\
    DDO 216 & (Pegasus)     &  -- & 7-10 & -- & -12.3 & $955 \pm 50$\\
    \tableline
  \end{tabular}
  {\footnotesize $^1$ This work}
\end{table}

The observations so far were carried out in $R$ and $B$-Band, sparsely
sampling a three year period starting with test observations in 1999.
This part of the data set should be sensitive for long period variable
stars with periods up to $\sim 500$ days. Additional observations in
$R$, $B$ and $I$-Band were obtained during 3 observing campaigns
at the 1.23~m telescope on Calar Alto densely sampling three two week
long periods. These observations should provide a ground for a search
for variable stars with shorter periods ranging from $\sim 1.5$ days
up to $\sim 10$ days.

The acquired data were bias subtracted, flat-fielded and cosmic ray
rejected. 
Then, the images from one night were astrometrically aligned to a
common reference frame and combined with individual weights
proportional to their $S/N$. For each epoch, consisting of all the
stacked images of a single night, a difference image against a common
deep reference frame was created using an implementation (G\"ossl \&
Riffeser, 2002, 2003) of the Alard algorithm (Alard \& Lupton, 1998).
Finally, these difference images were convolved with a stellar PSF.

To extract lightcurves from the reduced data, first all pixels
deviating significantly ($3\sigma$) from the reference image in a
minimum number of epochs $n$ were flagged, utilizing the complete
per-pixel error propagation of our data reduction pipeline. Then,
using these coordinates as input, values and associated errors are
read from the difference images and the lightcurve data are assembled.
To search for periodic signals in the extracted difference fluxes, a
Lomb (1976) algorithm using the interpretation from Scargle (1982) is
applied.

The photometric calibration was conducted using the HST data published
by Schulte-Ladbeck et al. (2003).

\section{Preliminary Results}

For the galaxies Leo~A, and UGCA~92, we have a very good monitoring
and a large fraction of the data passed already the pipeline. The
Leo~A data set serves as test case: 
A total of 26 variable star candidates were detected. 
Among them, we identified 16 secure long period variables (typical
average values $19.4 < R < 22.1$, and $74<$ period [days] $< 590$),
and we have 8 further candidates for LPVs.  In addition we were able
to identify two good candidates for $\delta$ Cephei stars with best
fitting periods of 6.4 and 1.69 days. The later candidate was
previously described by Dolphin et al. (2002) as C2-V58 with a period
of 1.4 days. The Dolphin et al. period solution fails in deriving a
reliable lightcurve with our data, yet, applying our period value to
their data set yields reasonable results.  The phase convolved
lightcurves for the two $\delta$ Cephei variables are shown in
Figure~1. 

\begin{figure}[t]
  \plottwo{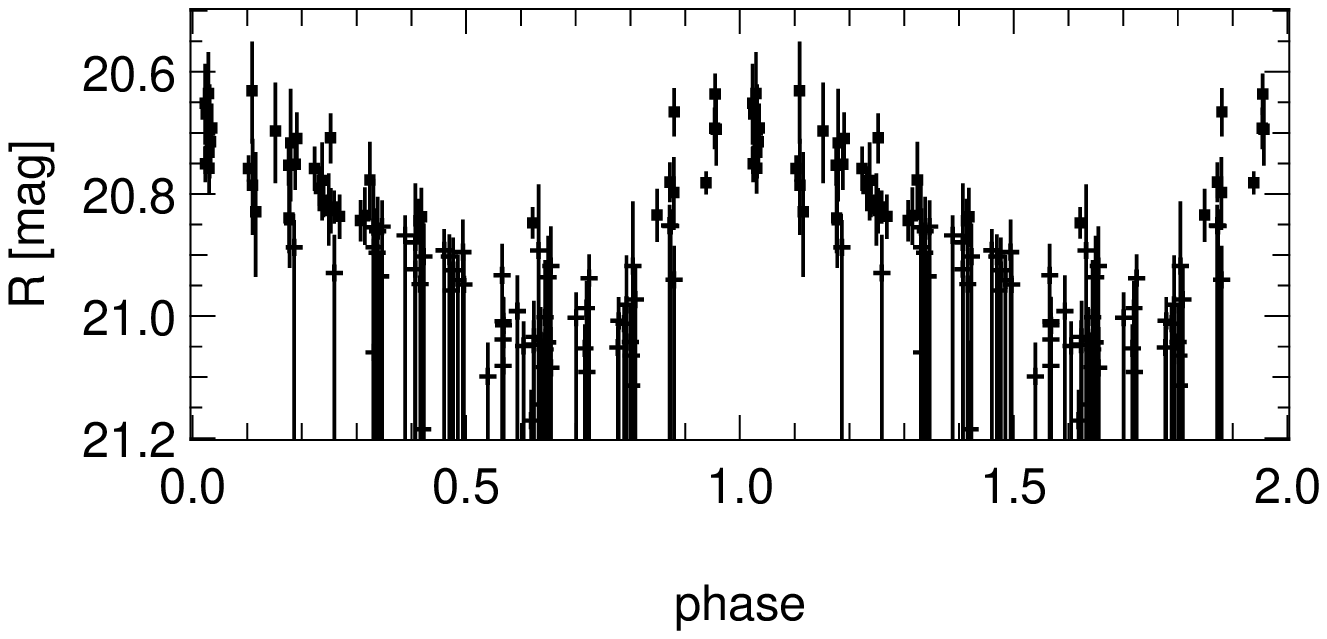}{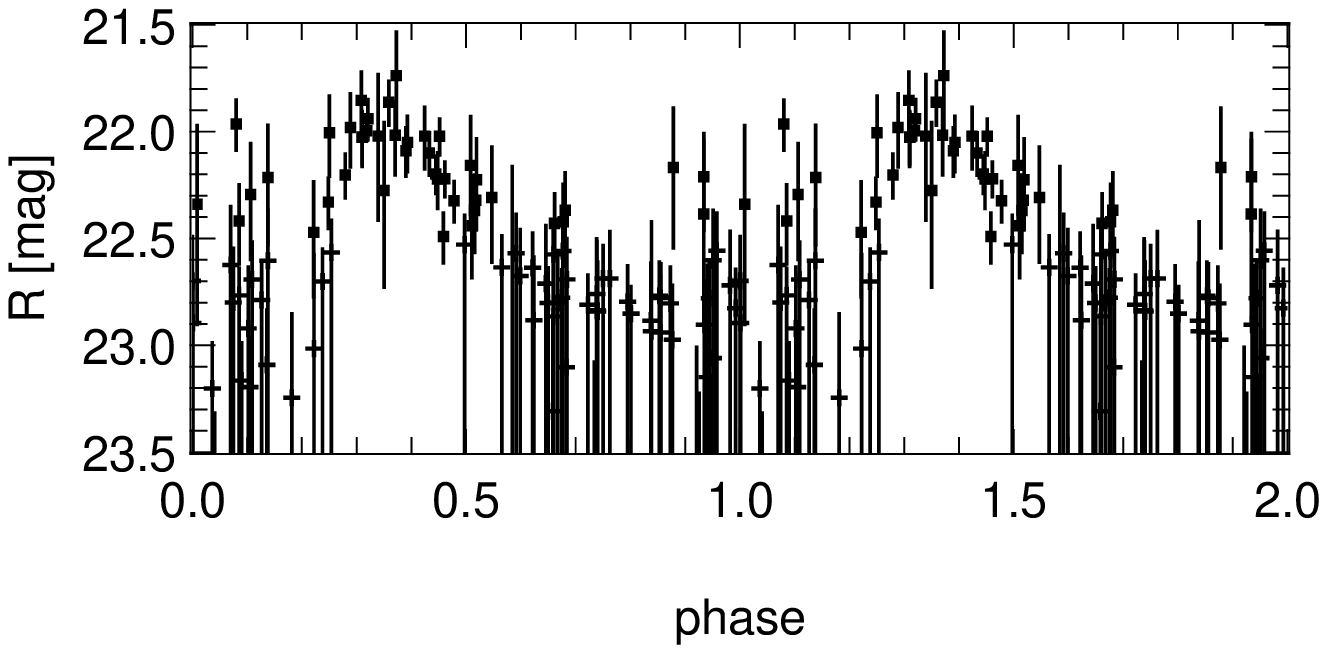}
  \caption{Phase convolved lightcurves for two Cepheids found in the
    Leo A dwarf galaxy. Plotted are the apparent $R$-Band magnitudes
    against twice the phase. The left panel shows a Cepheid with 6.4
    days period. In the right panel a Cepheid with a period of 1.69
    days is shown, that was previously published by Dolphin et
    al. (2002) with a period of 1.4 days.}
  \label{fig:ceph2}
\end{figure}

\section{Comparison with published work}

The color magnitude diagram shown in the left panel of Figure~2 is
based upon the HST data published by Tolstoy et al. (1996) and
Schulte-Ladbeck et al.  Flagged by bigger symbols are those variables
from our sample that lie inside the HST field of view, two $\delta$
Cephei variables in the instability strip (crosses) and the candidates
for long term variability (triangles) in the regime of the red giants.

\begin{figure}[t]
  \plottwo{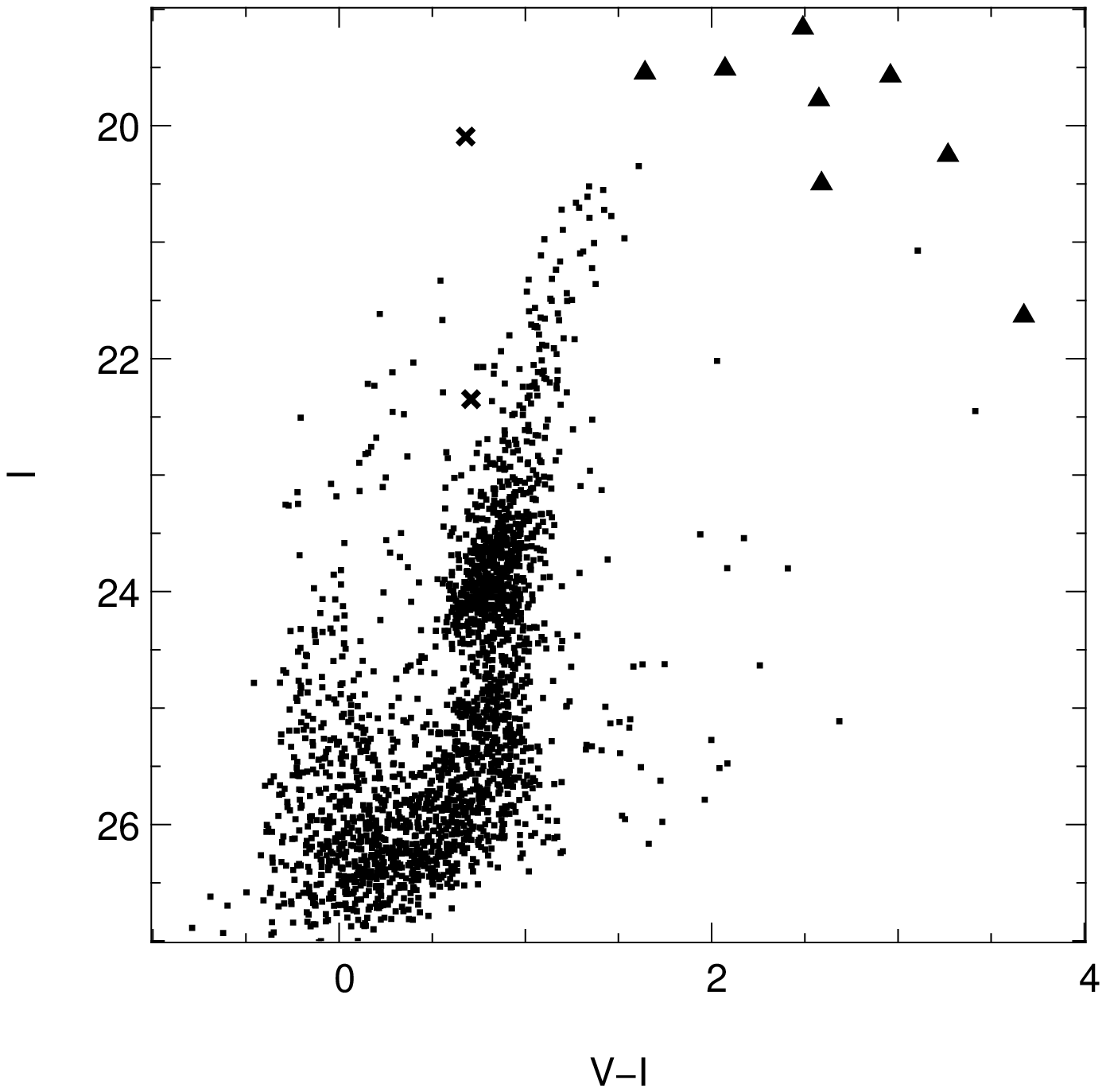}{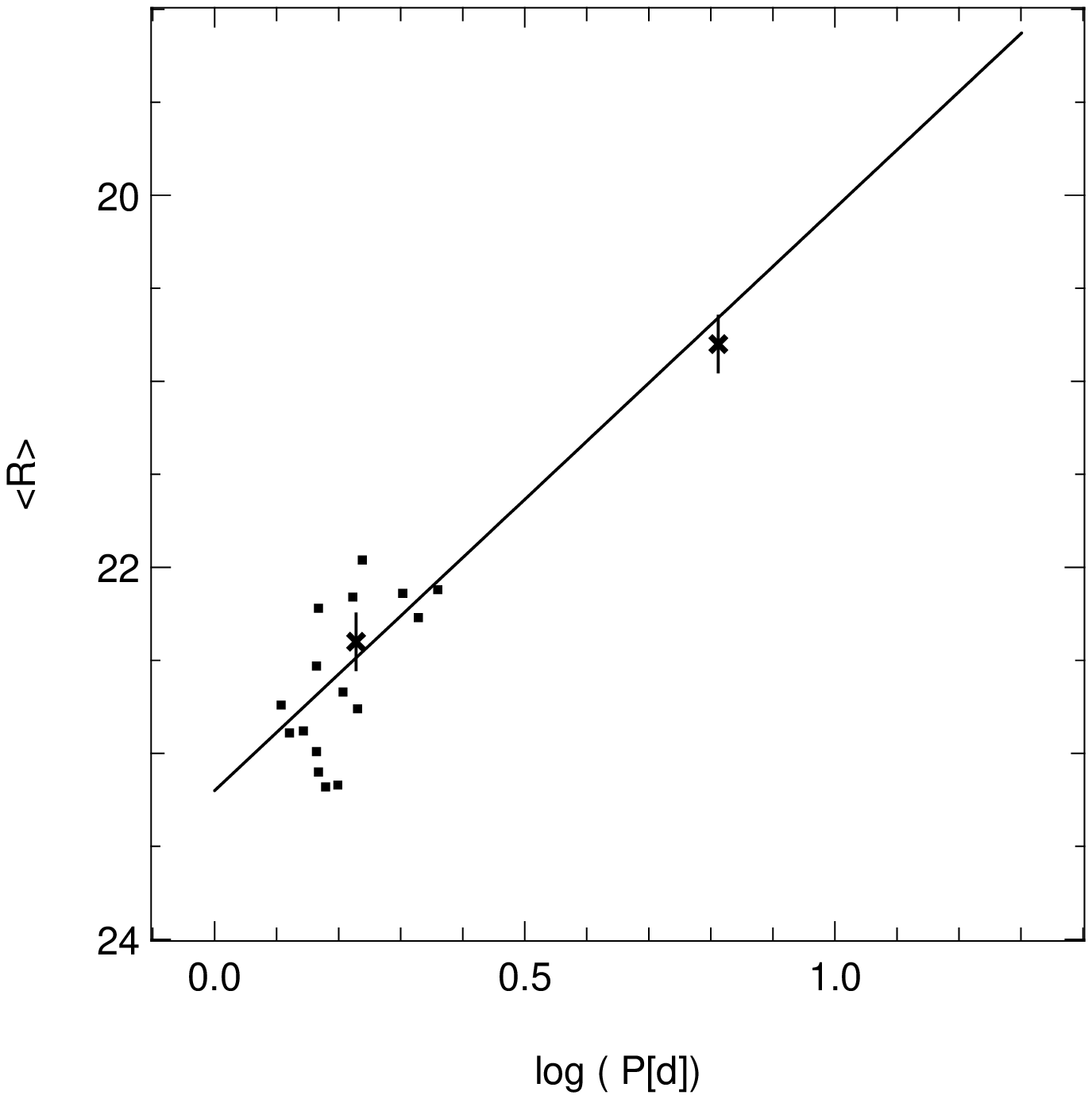}
  \caption{The left hand panel shows a color-magnitude diagram based
    on HST observations by Schulte-Ladbeck et al. (2003).  Overplotted
    the secured variables in Leo A with positions falling into the
    field of view of the HST observations. The right hand panel shows
    the period-luminosity relation of the SMC shifted to the distance
    determined by Tolstoy et al. (1996). The small dots mark the
    Cepheids found by Dolphin et al. (2002), the large crosses the two
    Cepheids presented here. 
  }
  \label{fig:pl-colmag}
\end{figure}

Tolstoy et al. (1996) based on ground-based data found a distance
modulus for Leo~A of 24.2 and a resulting distance of 690 kpc (see
also Schulte-Ladbeck et al.). This result got further support by the
search for short periodic variables with the WIYN telescope within 3
consecutive days in Dec. 2000 (Dolphin et al. 2002). Our data
complement this dataset for longer periods.

The right hand panel of Figure~2 shows the period-luminosity (PL)
relation of the SMC shifted to the distance determined by Tolstoy et
al. The short period variables measured by Dolphin coincide with the
shown PL relation. The overplotted values for the two Cepheids from
our survey (crosses) support this relation also in the regime of
longer periods.

\section{Summary}

We presented preliminary results for our survey for variable stars in
a sample of irregular local group dwarf galaxies. For the Leo~A dwarf
galaxy, the best analysed case so far, we already identified a total of
26 candidates for variability, 16 of these as long period variables
and 2 $\delta$ Cephei stars. We compared the later with the
period-luminosity relation and the short period variables discussed by
Dolphin et al. (2002). We found, that our Cepheids fully support their
findings and the resulting distance estimate for Leo~A. This result is
further in good agreement with the TRGB distance (Tolstoy et al.,
Schulte-Ladbeck et al.). The location of the LPVs in the
color-magnitude diagram indicate that most of them are early
asymptotic giant branch stars. While a complete census of these
intermediate age stars is missing for most of the Local Group members,
a proper statistic of their appearance can guide the reconstruction of
the star formation history at the age of several Gyr by-passing the
age metalicity degeneracy inherent to color magnitude diagram studies.

\acknowledgements We like to thank Drs. I. Drozdovsky, C. Maraston,
R.E. Schulte-Lad\-beck, and E. Tolstoy for helpful discussion. We
acknowledge the support of the Calar Alto and Wendelstein staff. J.
Fliri and A. Riffeser carried out some of our observations. The project is
supported by the Deutsche Forschungsgemeinschaft grant Ho~1812/3-1 and
Ho~1812/3-2.

\end{document}